\documentclass[%
 reprint,
 amsmath,amssymb,
 aps,
]{revtex4-1}

\usepackage{graphicx}
\usepackage{dcolumn}
\usepackage{bm}

\usepackage{graphicx,bm,color,amsmath,amssymb,amstext,amsfonts,caption,float,comment,hyperref}

\newcommand{\ba}{\begin{eqnarray}}
\newcommand{\ea}{\end{eqnarray}}
\newcommand{\be}{\begin{equation}}
\newcommand{\ee}{\end{equation}}
\newcommand{\bi}{\begin{itemize}}
\newcommand{\ei}{\end{itemize}}

\newcommand{\al}{\alpha}
\newcommand{\bt}{\beta}
\newcommand{\ga}{\gamma}

\newcommand{\la}{\lambda}

\newcommand{\cM}{{\cal M}}

\newcommand{\LF}{\left(}
\newcommand{\RF}{\right)}
\newcommand{\LT}{\left[}
\newcommand{\RT}{\right]}
\newcommand{\Ld}{\left.}
\newcommand{\Rd}{\right.}

\newcommand{\kb}{\bar{k}}

\newcommand{\mt}{\mathtt}

\newcommand{\non}{\nonumber\\}

\newcommand{\pb}{\bar{p}}

\begin{document}

\title{Scale of non-locality for a system of $n$ particles}

\author{Spyridon Talaganis}
\email{s.talaganis@lancaster.ac.uk}
\author{Ali Teimouri}
 \email{a.teimouri@lancaster.ac.uk}

\affiliation{Consortium for Fundamental Physics, Lancaster University, Lancaster, LA1 4YB, United Kingdom}


\begin{abstract}
Higher derivative theories of gravity are associated with a mass scale to insure the correct dimensionality of the covariant derivatives. This mass scale is known as the scale of non-locality. In this paper, by considering a higher derivative toy model,  we show that for a system of $n$ particles the effective mass scale is inversely proportional to the square root of the number of particles. We demonstrate that as the number of particles increases the corresponding effective mass scale associated with the  scattering amplitude decreases. 
\end{abstract}

\pacs{Valid PACS appear here}
\maketitle


\section{\label{sec:level1}Introduction}

Modified theories of gravity are expedient approach to resolve some of the problems of the Einstein theory of general relativity \cite{Clifton:2011jh}. Modifying general theory of relativity can be done in numerous ways. For instance,  it is possible to generalise the Einstein Hilbert action by writing a power series for the scalar curvature. This approach is known as $f(R)$ gravity \cite{Sotiriou:2008rp}. It is also natural to add higher order curvatures to the gravitational action, examples of this modification are best known by Lovelock gravity \cite{Lovelock:1971yv} or Gauss-Bonnet gravity, which is the simplified version of the Lovelock gravity. In addition to these modifications, one can consider higher derivative actions. In these cases one can act covariant derivatives on the curvature. 

Recently, inspired by string field theory, \cite{Biswas:2005qr} proposed a gravitational action where the Einstein-Hilbert action is modified by adding  higher order terms where infinite number of  covariant derivatives act on scalar curvature, Ricci and Riemann tensors. Such type of action is allowed by general covariance. It has been shown that such action provides interesting features; for instance such theory is shown to be ghost-free, or that it prevents singularity. Different aspects of this theory  were studied in \cite{Biswas:2011ar,Biswas:2013cha,Conroy:2015nva,Teimouri:2016ulk,Tal}.

Inspired by such gravitational theory, it is always possible to construct a toy model where  the action contain scalar fields rather than Riemannian tensors. In this case, we are going to keep the covariant derivatives and operate them on the scalar fields. Such approach simplifies the action essentially and hence studying the properties of the higher derivative theories would be easier. Some of the features of such toy model were explored extensively in \cite{Talaganis:2014ida,Talaganis:2016ovm,Talaganis:2017tnr}.

Scattering amplitude plays a crucial rule in quantum field theory (QFT). It is possible to obtain the S-matrix from the scattering amplitude and extract information about  probability amplitude  and the cross sections of various interactions \cite{Giddings:2011xs}. Moreover, in the limit of large center of mass energy
one can use eikonal approximation to obtain the behaviour of scattering amplitude \cite{Kabat:1992tb}. Similar analysis was done in \cite{Talaganis:2016ovm} to show that in the context of infinite derivative.

Previously it has been shown that,  in a scattering event, as the number of particles increases the scattering amplitude becomes more exponentially suppressed \cite{Talaganis:2016ovm}. In this paper we demonstrate that the amplitude is associated to an effective mass scale. The effective mass scale $M_{eff}$ satisfies, for large $n$, $M_{eff}\sim M/\sqrt{n}$, where $M$ is the mass scale (\textit{i.e.} scale of non-locality) and $n$ is the number of particles.

\section{Scale of Non-Locality}

We want to find the scale of non-locality for a system of $n$ particles.
Let us consider the scalar toy model
\be\label{sid}
S=S_{\mt{free}}+S_{\mt{int}}\,,
\ee
where
\be \label{free}
S_{\mt{free}}=\frac{1}{2}\int d^4 x \, \LF  \phi \Box a(\Box) \phi\RF
\ee
and 
\be
S_{\mt{int}}=\la \int d^4 x \, \LF \phi \Box \phi a(\Box) \phi \RF \,.
\ee
We choose
\be \label{eq:exp}
a (\Box) = e^{- \Box / M ^ {2}}\,,
\ee
where $M$ is the mass scale at which the non-local modifications become important.
The propagator in momentum space is then given in Euclidean space by
\be
\Pi (k ^ 2)= \frac{- i}{k^2 e ^ {\kb ^ 2}}\,,
\ee
where barred $4$-momentum vectors denote $\bar k = k/M$. The vertex factor
for the three incoming momenta $k_{1},~k_{2},~k_{3},$ which satisfies the
following conservation law,
\be
k _ {1} + k _ {2} + k _ {3} = 0\,,
\ee
is given by
\begin{align} \label{eq:uaua}
\la V (k _ {1}, k _ {2}, k _ {3}) & = -i \la \LT k_{1}^{2}(e^{\kb_{2}^2}+e^{\kb_{3}^2})+
k_{2}^{2}(e^{\kb_{3}^2}\Rd \non
& \Ld +  e^{\kb_{1}^2})+ k_{3}^{2}(e^{\kb_{1}^2}+e^{\kb_{2}^2}) \RT \,.
\end{align}
\begin{figure}[t]
\centering
\includegraphics[width=.40\textwidth]{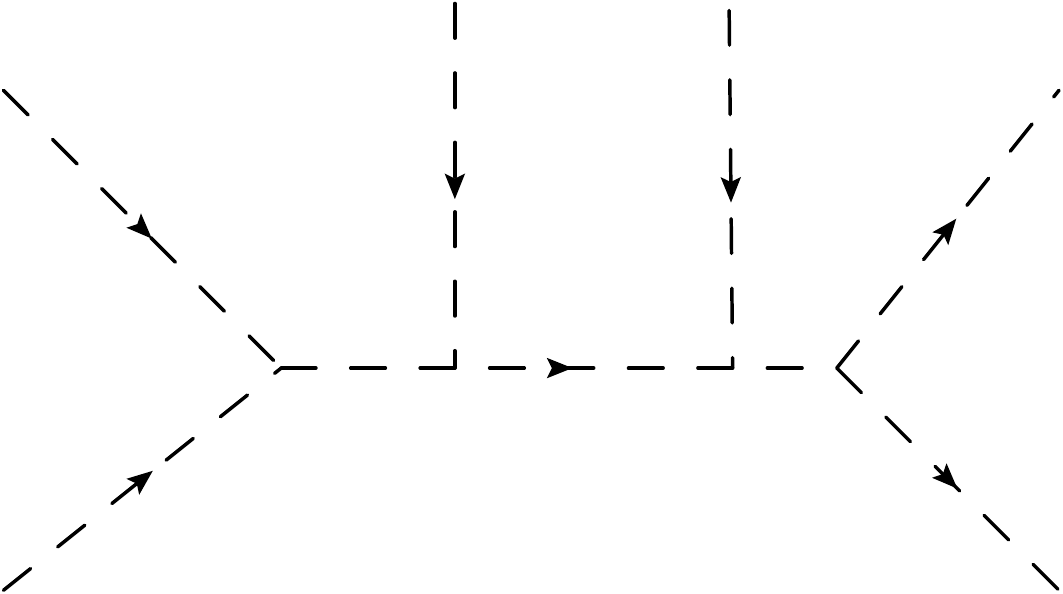}
\caption{\label{fig:tree} {\small The tree-level $6$-point scattering diagram.
The external momenta in the middle are $p_{5}$ and $p_{6}$.}}
\end{figure}
We shall note that at large momenta the dressed propagator goes as \cite{Talaganis:2016ovm}
\begin{equation}
\tilde\Pi(k^{2})\approx(1+4\bar k^{-2})^{-1}\ e^{-\frac{3\bar k^{2}}{2}}.
\end{equation}
Suppose we have a tree-level $6$-point scattering amplitude. We have that
\begin{align}\label{6}
i\cM &= \la^4 V(p_{1},p_{2},-p_{1}-p_{2})V(p_{3},p_{4},p_{1}+p_{2}+p_{5}+p_{6})\non
& \times V(p_{1}+p_{2},p_{5},-p_{1}-p_{2}-p_{5})\non
& \times V(p_{1}+p_{2}+p_{5},p_{6},-p_{1}-p_{2}-p_{5}-p_{6}) \non
& \times \frac{i}{(p_{1}+p_{2})^2 e ^ {(\pb_{1}+\pb_{2}) ^ 2}}\frac{1}{(p_{1}+p_{2}+p_{5})^2
e ^ {(\pb_{1}+\pb_{2}+\pb_{5}) ^ 2}}\non
& \times \frac{1}{(p_{1}+p_{2}+p_{5}+p_{6})^2 e ^ {(\pb_{1}+\pb_{2}+\pb_{5}+\pb_{6})
^ 2}}\,.
\end{align}
We have, from conservation of momentum, that,
\be
p_{1}+p_{2}+p_{3}+p_{4}+p_{5}+p_{6}=0\,.
\ee

Suppose we have $n$ vertices in the tree-level diagram and we want to find
$M_{\text{eff}}$. 
From Eq.~\eqref{6}, we have,
\begin{align}\label{10}
\cM &= \la^{4} \left\{\LT p_{1}^{2}(e^{\pb_{2}^2}+e^{(\pb_{1}+\pb_{2})^2})+
p_{2}^{2}(e^{\pb_{1}^2}+e^{(\pb_{1}+\pb_{2})^2})\Rd \Rd \non
& \Ld+ (p_{1}+p_{2})^{2}(e^{\pb_{1}^2}+e^{\pb_{2}^2}) \RT  \non
& \times \LT p_{3}^{2}(e^{\pb_{4}^2}+e^{(\pb_{1}+\pb_{2}+\pb_{5}+\pb_{6})^2})+
p_{4}^{2}(e^{\pb_{3}^2}+e^{(\pb_{1}+\pb_{2}+\pb_{5}+\pb_{6})^2})\Rd \non
& \Ld + (p_{1}+p_{2}+p_{5}+p_{6})^{2}(e^{\pb_{3}^2}+e^{\pb_{4}^2}) \RT \non
& \times \LT p_{5}^{2}(e^{(\pb_{1}+\pb_{2})^2}+e^{(\pb_{1}+\pb_{2}+\pb_{5})^2})\Rd
\non &+ (p_{1}+p_{2})^{2}(e^{\pb_{5}^2}+e^{(\pb_{1}+\pb_{2}+\pb_{5})^2})\non
& \Ld + (p_{1}+p_{2}+p_{5})^{2}(e^{\pb_{5}^2}+e^{(\pb_{1}+\pb_{2})^2}) \RT
\non
& \times \LT p_{6}^{2}(e^{(\pb_{1}+\pb_{2}+\pb_{5})^2}+e^{(\pb_{1}+\pb_{2}+\pb_{5}+\pb_{6})^2})\Rd
\non
& \Ld + (p_{1}+p_{2}+p_{5})^{2}(e^{\pb_{6}^2}+e^{(\pb_{1}+\pb_{2}+\pb_{5}+\pb_{6})^2})\Rd
\non
& \Ld \Ld + (p_{1}+p_{2}+p_{5}+p_{6})^{2}(e^{\pb_{6}^2}+e^{(\pb_{1}+\pb_{2}+\pb_{5})^2})
\RT \right\} \non
& \times \frac{1}{(p_{1}+p_{2})^2 e ^ {(\pb_{1}+\pb_{2}) ^ 2}}\frac{1}{(p_{1}+p_{2}+p_{5})^2
e ^ {(\pb_{1}+\pb_{2}+\pb_{5}) ^ 2}}\non
& \times \frac{1}{(p_{1}+p_{2}+p_{5}+p_{6})^2 e ^ {(\pb_{1}+\pb_{2}+\pb_{5}+\pb_{6})
^ 2}} \,.
\end{align}

If we expand Eq.~\eqref{10}, we shall get terms of the form,
\be
\sum (\mathrm{polynomial~in~p})e^{\sum (\mathrm{polynomial~in~p})/M^{2}},
\ee
coming from the vertices.

Suppose now that we dress the four vertices. At sufficiently high loop order
$n$ (when $n\geqslant4$), the exponents in the dressed
vertices become negative. The vertex factors are:
\be \label{dara}
e^{\al^{n}\pb_{1}^{2}+\bt^{n}\pb_{2}^{2}+\ga^{n}\pb_{3}^{2}} \,,
\ee
where $p_{1}$, $p_{2}$, $p_{3}$ are the incoming vertex momenta. When the
loop order $n$ of the dressed vertices is equal to $4$, that is, $n=4$,
the exponents for the dressed vertices in Eq.~\eqref{dara} become negative~\cite{Talaganis:2014ida,Talaganis:2016ovm}:
\be
\al^{4}=\bt^{4}=\ga^{4}= -\frac{11}{27} \,.
\ee
Then, going to Euclidean space, we have, for the largest external momentum
contribution, ($n=4$),
\begin{align}
\mathcal{M}^{'} & \sim e^{-\frac{11\pb_{1}^{2}}{27}}e^{-\frac{11\pb_{2}^{2}}{27}}e^{-\frac{11\pb_{3}^{2}}{27}}e^{-\frac{11\pb_{4}^{2}}{27}}e^{-\frac{11\pb_{5}^{2}}{27}}e^{-\frac{11\pb_{6}^{2}}{27}}
\non
& \times e^{-\frac{22(p_{1}+p_{2})^2}{27M^2}}e^{-\frac{22(p_{1}+p_{2}+p_{5})^2}{27M^2}}e^{-\frac{22(p_{1}+p_{2}+p_{5}+p_{6})^2}{27M^2}}
\non
& \times e^{-\frac{(p_{1}+p_{2})^2}{M^2}}e^{-\frac{(p_{1}+p_{2}+p_{5})^2}{M^2}}e^{-\frac{(p_{1}+p_{2}+p_{5}+p_{6})^2}{M^2}}\non
& = e^{-\frac{11\pb_{1}^{2}}{27}}e^{-\frac{11\pb_{2}^{2}}{27}}e^{-\frac{11\pb_{3}^{2}}{27}}e^{-\frac{11\pb_{4}^{2}}{27}}e^{-\frac{11\pb_{5}^{2}}{27}}e^{-\frac{11\pb_{6}^{2}}{27}}
e^{-\frac{49(p_{1}+p_{2})^2}{27M^2}}\non
& \times e^{-\frac{49(p_{1}+p_{2}+p_{5})^2}{27M^2}}e^{-\frac{49(p_{1}+p_{2}+p_{5}+p_{6})^2}{27M^2}}
 \,.
\end{align}

If $\lvert p_{1} \rvert =\lvert p_{2} \rvert =\lvert p_{3} \rvert =\lvert
p_{4} \rvert =\lvert p_{5} \rvert =\lvert p_{6} \rvert =\lvert p \rvert$
and $p_1=p_3=p_5=p$, $p_2=p_4=p_6=-p$, then
\be
\mathcal{M}^{'} \sim e^{-\frac{115 \pb^{2}}{27}} \,.
\ee

Now suppose we have an $8$-point tree-level scattering diagram. Then we have,
for the largest external momentum contribution,  ($n=4$),
\begin{align}
\mathcal{M}^{'} & \sim e^{-\frac{11\pb_{1}^{2}}{27}}e^{-\frac{11\pb_{2}^{2}}{27}}e^{-\frac{11\pb_{3}^{2}}{27}}e^{-\frac{11\pb_{4}^{2}}{27}}e^{-\frac{11\pb_{5}^{2}}{27}}e^{-\frac{11\pb_{6}^{2}}{27}}
e^{-\frac{11\pb_{7}^{2}}{27}}e^{-\frac{11\pb_{8}^{2}}{27}}\non
& \times e^{-\frac{22(p_{1}+p_{2})^2}{27M^2}}e^{-\frac{22(p_{1}+p_{2}+p_{5})^2}{27M^2}}e^{-\frac{22(p_{1}+p_{2}+p_{5}+p_{6})^2}{27M^2}}
\non
& \times e^{-\frac{22(p_{1}+p_{2}+p_{5}+p_{6}+p_{7})^2}{27M^2}}e^{-\frac{22(p_{1}+p_{2}+p_{5}+p_{6}+p_{7}+p_{8})^2}{27M^2}}\non
& \times e^{-\frac{(p_{1}+p_{2})^2}{M^2}}e^{-\frac{(p_{1}+p_{2}+p_{5})^2}{M^2}}e^{-\frac{(p_{1}+p_{2}+p_{5}+p_{6})^2}{M^2}}\non
& \times e^{-\frac{(p_{1}+p_{2}+p_{5}+p_{6}+p_{7})^2}{M^2}}e^{-\frac{(p_{1}+p_{2}+p_{5}+p_{6}+p_{7}+p_{8})^2}{M^2}}\non
& = e^{-\frac{11\pb_{1}^{2}}{27}}e^{-\frac{11\pb_{2}^{2}}{27}}e^{-\frac{11\pb_{3}^{2}}{27}}e^{-\frac{11\pb_{4}^{2}}{27}}e^{-\frac{11\pb_{5}^{2}}{27}}e^{-\frac{11\pb_{6}^{2}}{27}}e^{-\frac{11\pb_{7}^{2}}{27}}e^{-\frac{11\pb_{8}^{2}}{27}}\non
& \times e^{-\frac{49(p_{1}+p_{2})^2}{27M^2}}e^{-\frac{49(p_{1}+p_{2}+p_{5})^2}{27M^2}}e^{-\frac{49(p_{1}+p_{2}+p_{5}+p_{6})^2}{27M^2}}\non
& \times e^{-\frac{49(p_{1}+p_{2}+p_{5}+p_{6}+p_{7})^2}{27M^2}}e^{-\frac{49(p_{1}+p_{2}+p_{5}+p_{6}+p_{7}+p_{8})^2}{27M^2}}
 \,.
\end{align}
Again, from the conservation of momentum: 
\be
p_{1}+p_{2}+p_{3}+p_{4}+p_{5}+p_{6}+p_{7}+p_{8}=0\,.
\ee

If $\lvert p_{1} \rvert =\lvert p_{2} \rvert =\lvert p_{3} \rvert =\lvert
p_{4} \rvert =\lvert p_{5} \rvert =\lvert p_{6} \rvert=\lvert p_{7} \rvert=\lvert
p_{8} \rvert = \lvert p \rvert$ and $p_1=p_3=p_5=p_7=p$, $p_2=p_4=p_6=p_{8}=-p$,
then
\be
\mathcal{M}^{'} \sim e^{-\frac{186 \pb^{2}}{27}}=e^{-\frac{62 \pb^{2}}{9}}
\,.
\ee
We observe that the $8$-point diagram is even more strongly exponentially
suppressed in the UV as compared to the $6$-point diagram.

For a $2n$-point tree-level diagram with dressed vertices, where $\lvert
p_{i} \rvert=p$, $i=1,\dots,2n$, $p_{2j-1}=p$, $p_{2j}=-p$, $j=1,\dots,n$,
we have ($n \geq 2$)
\be
\mathcal{M}^{'} \sim e^{-\frac{(22n+49(n-2)) \pb^{2}}{27}}= e^{-\frac{(71n-98)
\pb^{2}}{27}} \,.
\ee
We can write the equation above as:
\be
\cM^{'} \sim e ^{- \LF \frac{p}{M_{eff}} \RF ^2} \,,
\ee
where
\be
M_{eff} = \LF \frac{27}{71n-98}\RF^{1/2} M \,.
\ee

\subsection{External momenta on the external legs}

Suppose we have external momenta on the external legs in a $6$-point, tree-level
diagram (one on one of the legs on the left-hand side of the diagram and
the other on one of the legs on the right-hand side of the diagram). Now
we will compute the amplitude $\cM$. Employing dressed vertices, we obtain:
\be
\cM \sim e^{-\frac{115\pb^2}{27}} \,.
\ee

\subsection{Dressing the vertices with $1$-loop diagram in the middle}

A $1$-loop diagram with external momenta $p$, $-p$ goes as $e^{3\pb^{2}/2}$
for large external momenta. Adding a $1$-loop diagram in the middle in Fig.~\ref{fig:tree},
where the propagators and the vertices are both dressed, gives us a scattering
amplitude that goes to zero for large external momenta. Thus:\begin{align}
\mathcal{M}^{'} & \sim e^{-\frac{11\pb_{1}^{2}}{27}}e^{-\frac{11\pb_{2}^{2}}{27}}e^{-\frac{11\pb_{3}^{2}}{27}}e^{-\frac{11\pb_{4}^{2}}{27}}e^{-\frac{11\pb_{5}^{2}}{27}}e^{-\frac{11\pb_{6}^{2}}{27}}
e^{-\frac{22(p_{1}+p_{2})^2}{27M^2}}\non
& \times e^{-\frac{22(p_{1}+p_{2}+p_{5})^2}{27M^2}}e^{-\frac{22(p_{1}+p_{2}+p_{5}+p_{6})^2}{27M^2}}
\non
& \times e^{-\frac{(p_{1}+p_{2})^2}{M^2}}e^{-\frac{2(p_{1}+p_{2}+p_{5})^2}{M^2}}e^{\frac{3(p_{1}+p_{2}+p_{5})^2}{2M^2}}e^{-\frac{(p_{1}+p_{2}+p_{5}+p_{6})^2}{M^2}}\non
& = e^{-\frac{11\pb_{1}^{2}}{27}}e^{-\frac{11\pb_{2}^{2}}{27}}e^{-\frac{11\pb_{3}^{2}}{27}}e^{-\frac{11\pb_{4}^{2}}{27}}e^{-\frac{11\pb_{5}^{2}}{27}}e^{-\frac{11\pb_{6}^{2}}{27}}
e^{-\frac{49(p_{1}+p_{2})^2}{27M^2}}\non
& \times e^{-\frac{71(p_{1}+p_{2}+p_{5})^2}{54M^2}}e^{-\frac{49(p_{1}+p_{2}+p_{5}+p_{6})^2}{27M^2}}
 \,.
\end{align}
If $\lvert p_{1} \rvert =\lvert p_{2} \rvert =\lvert p_{3} \rvert =\lvert
p_{4} \rvert =\lvert p_{5} \rvert =\lvert p_{6} \rvert =\lvert p \rvert$
and $p_1=p_3=p_5=p$, $p_2=p_4=p_6=-p$, then,
\be
\mathcal{M}^{'} \sim e^{-\frac{203 \pb^{2}}{54}} \,.
\ee
For a $2n$-point tree-level diagram with dressed vertices, where $\lvert
p_{i} \rvert=p$, $i=1,\dots,2n$, $p_{2j-1}=p$, $p_{2j}=-p$, $j=1,\dots,n$,
we have ($n \geq 2$)
\be
\mathcal{M}^{'} \sim e^{-\frac{(44n+98(n-2)-27) \pb^{2}}{54}}= e^{-\frac{(142n-223)
\pb^{2}}{54}} \,.
\ee
We can write the equation above as
\be
\cM^{'} \sim e ^{- \LF \frac{p}{M_{eff}} \RF ^2} \,,
\ee
where
\be
M_{eff} = \LF \frac{54}{142n-223}\RF^{1/2} M \,.
\ee

\subsection{Dressing both the propagators and the vertices}

Suppose we dress both the propagators and the vertices in Fig~\ref{fig:tree}.
We wish to find the behaviour of the scattering amplitude $\cM^{'}$for large
external momenta. We have,
\begin{align}
\mathcal{M}^{'} & \sim e^{-\frac{11\pb_{1}^{2}}{27}}e^{-\frac{11\pb_{2}^{2}}{27}}e^{-\frac{11\pb_{3}^{2}}{27}}e^{-\frac{11\pb_{4}^{2}}{27}}e^{-\frac{11\pb_{5}^{2}}{27}}e^{-\frac{11\pb_{6}^{2}}{27}}
e^{-\frac{22(p_{1}+p_{2})^2}{27M^2}}\non
& \times e^{-\frac{22(p_{1}+p_{2}+p_{5})^2}{27M^2}}e^{-\frac{22(p_{1}+p_{2}+p_{5}+p_{6})^2}{27M^2}}
\non
& \times e^{-\frac{3(p_{1}+p_{2})^2}{2M^2}}e^{-\frac{3(p_{1}+p_{2}+p_{5})^2}{2M^2}}e^{-\frac{3(p_{1}+p_{2}+p_{5}+p_{6})^2}{2M^2}}\non
& = e^{-\frac{11\pb_{1}^{2}}{27}}e^{-\frac{11\pb_{2}^{2}}{27}}e^{-\frac{11\pb_{3}^{2}}{27}}e^{-\frac{11\pb_{4}^{2}}{27}}e^{-\frac{11\pb_{5}^{2}}{27}}e^{-\frac{11\pb_{6}^{2}}{27}}
e^{-\frac{125(p_{1}+p_{2})^2}{54M^2}}\non
& \times e^{-\frac{125(p_{1}+p_{2}+p_{5})^2}{54M^2}}e^{-\frac{125(p_{1}+p_{2}+p_{5}+p_{6})^2}{54M^2}}
 \,.
\end{align}
If $\lvert p_{1} \rvert =\lvert p_{2} \rvert =\lvert p_{3} \rvert =\lvert
p_{4} \rvert =\lvert p_{5} \rvert =\lvert p_{6} \rvert =\lvert p \rvert$
and $p_1=p_3=p_5=p$, $p_2=p_4=p_6=-p$, then
\be
\mathcal{M}^{'} \sim e^{-\frac{257 \pb^{2}}{54}} \,.
\ee
For a $2n$-point tree-level diagram with dressed vertices, where $\lvert
p_{i} \rvert=p$, $i=1,\dots,2n$, $p_{2j-1}=p$, $p_{2j}=-p$, $j=1,\dots,n$,
we have ($n \geq 2$)
\be \label{mmmm}
\mathcal{M}^{'} \sim e^{-\frac{(44n+125(n-2)) \pb^{2}}{54}}= e^{-\frac{(169n-250)
\pb^{2}}{54}} \,.
\ee
We can write the equation above as:
\be
\cM^{'} \sim e ^{- \LF \frac{p}{M_{eff}} \RF ^2} \,,
\ee
where
\be
M_{eff} = \LF \frac{54}{169n-250}\RF^{1/2} M \,.
\ee 

We observe that $M_{eff}$ decreases as the number of  particles $n$ increases. By dimensional analysis, one can write down the effective length scale $L_{eff}$ as $L_{eff}\sim M_{eff}^{-1}$. Hence, one can see that the effective length scale $L_{eff}$ increases as the number of particles $n$ increases.

\section{Conclusion}
In this paper we considered a scalar field toy model which is constructed by infinite derivatives and inspired by \textit{infinite derivative gravity}. It is possible to recast the infinite derivative function as $e^{- \Box / M^{2}}$, where we observe a mass scale due to the d'Alembertian operator's dimensionality. We have shown for such theory, there exists an effective mass scale which can be calculated by obtaining the relevant scattering amplitude.
This effective mass scale $M_{eff}$ is proportional to the mass scale of the theory (\textit{i.e.} scale of non-locality), $M,$ and it is also, for large $n$, inversely proportional to the square root of the number of incoming particles in a scattering event that is: $M_{eff}\sim M/\sqrt{n}$ to be precise. By dimensional analysis it is possible to relate the mass scale with the length scale (\textit{i.e.} $[L_{eff}]=[M_{eff}]^{-1}$). Since in string theory one would expect higher curvature term, namely infinite derivative contribution in curvature from $\alpha'$ corrections \cite{string}, it would be interesting to see if there is any possible relation between the effective length scale, $L_{eff}$, and string length (the natural length that appears in string theory), $l_s$ (we shall note that $\alpha'=l_s^2$). However, this requires one to apply the scattering analysis given in this paper straight to gravitational theory, which indeed is a challenging task.   
 \section*{Acknowledgement}              
S.T. is supported by a scholarship from the Onassis Foundation.

\end{document}